\documentclass[conference]{IEEEtran}
\IEEEoverridecommandlockouts
\usepackage[utf8]{inputenc} % For PDFLaTeX
\usepackage{cite}
\usepackage{subfiles}

\usepackage{float}
\usepackage{wrapfig}
\usepackage{caption}
\usepackage{subcaption}

\usepackage{amsmath,amssymb,amsfonts}
\usepackage{algorithmic}
\usepackage{graphicx}
\usepackage{textcomp}
\usepackage{xcolor}
\usepackage{tcolorbox}

\usepackage{rotating}
\def\BibTeX{{\rm B\kern-.05em{\sc i\kern-.025em b}\kern-.08em
    T\kern-.1667em\lower.7ex\hbox{E}\kern-.125emX}}

\makeatletter
\newcommand{\newlineauthors}{%
  \end{@IEEEauthorhalign}\hfill\mbox{}\par
  \mbox{}\hfill\begin{@IEEEauthorhalign}
}
\makeatother

\def\nLibs{24,285}   % number libs
\def\nJars{574,581}   % number jars
\def\nCVEs{1,285}   % number CVEs
\def\nHits{528,090}
\def\nCWEs{157}   % number CWEs

\title{CVE Breadcrumbs: Tracking Vulnerabilities Through Versioned Apache Libraries}
\author{
\IEEEauthorblockN{Derek Garcia}
\IEEEauthorblockA{\textit{Information and Computer Sciences} \\
\textit{University of Hawai'i}\\
Honolulu, USA \\
dgarcia2@hawaii.edu}
\and
\IEEEauthorblockN{Briana Lee}
\IEEEauthorblockA{\textit{Information and Computer Sciences} \\
\textit{University of Hawai'i}\\
Honolulu, USA \\
brianall@hawaii.edu}
\and
\IEEEauthorblockN{Ibrahim Matar}
\IEEEauthorblockA{\textit{Information and Computer Sciences} \\
\textit{University of Hawai'i}\\
Honolulu, USA \\
imatar77@hawaii.edu}
\newlineauthors
\IEEEauthorblockN{David Rickards}
\IEEEauthorblockA{\textit{Information and Computer Sciences} \\
\textit{University of Hawai'i}\\
Honolulu, USA \\
davidar@hawaii.edu}
\and
\IEEEauthorblockN{Andrew Zilnicki}
\IEEEauthorblockA{\textit{Information and Computer Sciences} \\
\textit{University of Hawai'i}\\
Honolulu, USA \\
zilnicki@hawaii.edu}
}

\begin{document}

\maketitle

\begin{abstract}
The Apache Software Foundation (ASF) ecosystem underpins a vast portion of modern software infrastructure, powering widely used components such as Log4j, Tomcat, and Struts. However, the ubiquity of these libraries has made them prime targets for high-impact security vulnerabilities, as illustrated by incidents like Log4Shell. Despite their widespread adoption, Apache projects are not immune to recurring and severe security weaknesses. We conduct a historical analysis of the Apache ecosystem to follow the "breadcrumb trail of vulnerabilities" by compiling a comprehensive dataset of Common Vulnerabilities and Exposures (CVEs) and Common Weakness Enumerations (CWEs). We examine trends in exploit recurrence, disclosure timelines, and remediation practices. Our analysis is guided by four key research questions: (1) What are the most persistent and repeated CWEs in Apache libraries? (2) How long do CVEs persist before being addressed? (3) What is the delay between CVE introduction and official disclosure? and (4) How long after disclosure are CVEs remediated? We present a detailed timeline of vulnerability lifecycle stages across Apache libraries and offer insights to improve secure coding practices, vulnerability monitoring, and remediation strategies. Our contributions include a curated dataset covering \nLibs\ Apache libraries, \nCVEs\ CVEs, and \nCWEs\ CWEs, along with empirical findings and developer-focused recommendations.
\end{abstract}

\section{Introduction}

% As part of your Goals section, elaborate on why this study is important and how it will benefit developers and software teams. What are they supposed to do with the outcomes/results of this work?There are some broken references --> [?]  

% Introduction:
% ○ Briefly introduce the area of research associated with your project and its
% importance for software development/maintenance
% ○ Present a compelling case for why this research area requires new solutions or
% perspectives. Providing actual real-world examples would help show the
% significance of the problem
% ○ Highlight the potential impact of addressing these challenges on software quality
% and maintenance practices
In today's software development landscape, the Apache Software Foundation (ASF) ecosystem has become a cornerstone of modern infrastructure. Apache projects such as Log4j, Tomcat, HTTP server, and OFBiz are used in numerous applications, cloud services, and countless other backend services. However, their ubiquity has made them an attractive and high-impact target for attackers.

One of the most well-known incidents - the Log4Shell \cite{nistCve202144228} shows how a single vulnerability in an Apache library can compromise thousands of systems globally. The remote code execution flaw in Apache Log4j exposed a wide range of services to exploitation due to its integration in numerous systems.

Other high-profile vulnerabilities in Apache reinforce this trend. For example, Apache HTTP Server's flaws could allow an attacker to perform smuggling attacks or bypass SSL client authentication\cite{nistCve202440725, nistCve202440898}. Apache Tomcat was found to be vulnerable to path equivalence attacks that allow unauthorized PUT requests\cite{nistCve202524813}.

These recurring and severe issues highlight a critical reality: Apache projects, despite their pedigree and widespread use, are not immune to security flaws. There is a need for focused research into how vulnerabilities evolve in the Apache ecosystem. Understanding patterns in vulnerability emergence, disclosure, and remediation can lead to actionable insights for both maintainers and the broader security community. By analyzing past Common Vulnerabilities and Exposures (CVEs) and Common Weakness Enumerations (CWEs) within the Apache ecosystem, it is possible to expose persistent flaws and evaluate the responsiveness of remediation efforts.

% Goal and Research Questions:
% ○ Clearly and concisely describe the issue you are trying to address
% ○ Describe how you envision your research supporting the developer community
% with building and maintaining their software systems
% ■ Discuss the immediate and long-term benefits
% ○ Specify the Research Questions (RQ) you intend to address in your study
% ○ For each RQ:
% ■ Explain how it explores previously unaddressed aspects of the problem
% ■ Demonstrate how it offers a novel perspective on existing solutions
% ■ Show how it feeds into the study's overall goal
% ○ Describe how answering these questions will make a unique contribution to the
% field Note: Phase III of the project will be answering these questions
% ○ Discuss how your work advances the state of the art in software maintenance
\subsection{Goal \& Research Questions}
Through compiling a comprehensive dataset, this study looks to provide a clear historical timeline of vulnerability management, disclosure, and remediation within the Apache ecosystem. This timeline can then be used to address prevalent risks in Apache libraries and identify gaps in securing these libraries. The insights gained from this analysis will inform the design of enhanced security strategies , the adoption of improved secure coding guidelines, or implementation of stronger vulnerability monitoring processes.  We aim to address these outcomes with the four following research questions (RQs):\\

\noindent\textbf{RQ1: What are the persistent and repeated exploits (CWEs) that appear throughout versions?}

\noindent By identifying exploit trends specific to the Apache ecosystems, we aim to highlight repeated exploits that must be addressed and should warrant greater attention in secure coding practices and organizational security oversight.  Historical trends of exploits can also be analyzed to correlate the reduction or rise of certain exploits with other major events in software development that can be used to draw parallels with potential future exploit trends. \\

\noindent\textbf{RQ2: How long do CVEs persist within libraries in the Apache ecosystem?}

\noindent The \textbf{CVE Presence Time} will highlight the average time vulnerabilities are contained in Apache library. We aim to address whether this time has been increasing or decreasing in recent years. This insight can be used to address  the overall state of vulnerability management within Apache libraries.\\

\noindent\textbf{RQ3: How long is the duration between the first instance of a CVE in an Apache library to its official disclosure in NVD?}

\noindent The \textbf{CVE Disclosure Time} will address the rate at which vulnerabilities are discovered in Apache libraries and how proactive vulnerabilities are being reported.\\

\noindent\textbf{RQ4: How long is the duration between the official disclosure of a CVE in NVD and a released version of an Apache library where the CVE has been remediated?}

\noindent The \textbf{CVE Remediation Time} will address the rate at which vulnerabilities are patched in Apache libraries and how proactive vulnerabilities are being patched in software.\\

In this paper, our contributions are three-fold:
\begin{itemize}
    \item An aggregated dataset of \nLibs\ Apache libraries with \nCVEs\ related CVEs. and \nCWEs\ related CWEs
    \item A detailed timeline of CVE presence, disclosure, and remediation within Apache libraries.
    \item Key trends in exploit patterns and recommendations to developers.
\end{itemize}

% The paper is organized as follows: section IV presents the our detailed approach in collecting vulnerability timelines, section V 
\section{Background and Related Work}
% ○ Identify key gaps or limitations in current approaches that motivate your research
This section outlines the background surrounding historical vulnerability tracking in datasets and similar work.

\subsection{Malicious Package Datasets}
PyPi, NPM, Maven Central, and other major language repositories are critical for facilitating the growth and use of open-source code. To promote these rapid iterative development ecosystems, these registries often have very few bars to entry, often requiring little more than an email and username. Consequently, malicious packages can easily be uploaded to these registries and pollute the software supply chain. Previous work has aggregated these datasets in collections of malware and found techniques such as typosquatting, an obfuscation technique where a malicious package uses a similarly sounding name to a legitimate package, such as "nunpy" instead of "numpy"\cite{maurice_backstabbers_2020,guo_empirical_2023}. These datasets mainly focus on the PyPi and NPM ecosystems, leaving a gap in Java repositories such as Maven Central. This work also emphasizes a focus on malware rather than CVEs, with the distinction being malware is intentionally designed to be harmful. In contrast, CVEs originate from well-intentioned but either poorly designed or non-malicious mistakes in the coding process. Additionally, malware frequently relies on CVEs to perform exploits and by developing a timeline of CVEs among testing libraries, we will be able to identify attack surfaces not exposed in previous collections of purely malicious packages.

\subsection{CWE Trend Analysis} %title to change
A study analyzed the top 10 CWEs from 2011 to 2020, revealing persistent software weaknesses. The list included 19 of 88 possible-based, variant, and compound (BVC) CWEs and 17 of 39 pillars and class (PC) layer CWEs. The research in this study also found that injection and memory errors were the two primary weaknesses. SQL injection consistently ranked the most dangerous, with an average most significant security weaknesses (MSSW) score of 76.68. The findings emphasize the need for improved security measures as the common weaknesses persist in software systems today \cite{gueye_decade_2021}. This study calls for new approaches to identify persistent weaknesses and security practices to mitigate continuous weaknesses effectively. 

\subsection{CVE Trend Analysis}
Previous work has made extensive use of NVD as a trusted source of vulnerability data, in particular utilizing the components of the Common Vulnerability Scoring System (CVSS)\cite{noauthor_common_nodate}. This work found certain vulnerabilities and weaknesses were consistent throughout the period of time between 2005 to 2019 by focusing on all CVEs this period\cite{gueye_historical_2021}. However, CVSS has been subject to much criticism throughout its existence\cite{noauthor_software_nodate} and more modern vulnerability scoring systems such as Stakeholder-Specific Vulnerability Categorization (SSVC)\cite{cybersecurity_and_infrastructure_security_agency_cisa_2022} and Exploit Prediction Scoring System (EPSS)\cite{epss} have been developed to help address and improve vulnerability scoring. Additional work utilizing historical NVD data has also focused on Free and Open Source Software (FOSS)\cite{alexopoulos_how_2022}, utilizing eleven large FOSS project to identify the average lifespan of CVEs in software. Our approach aims to focus on a specific domain to isolate key threats specific to testing libraries to create a more relevant insight for software developers, while curating a large enough dataset to provide significant findings.
\section{Methodology}
This section will detail our approach to selecting ground truth sources to construct our own specialized dataset relative to the Apache ecosystem for our analysis.

\subsection{Data Sources}
When constructing our dataset, we aggregated several existing and reputable sources into a single dataset.

\vspace{.5em}
\noindent\textcircled{1} \textbf{Maven Central Repository}\\
The Maven Central Repository is a community-led repository where users can publish their libraries and Java artifacts. Initially developed as part of the Apache Maven build tool in the early 2000s, today it hosts millions of Java artifacts and is utilized by modern build tools such as Gradle, Apache Ivy, and SBT%\cite{}
. 

Crucially, the Maven Central Repository contains versioned releases with timestamps for hundreds of projects, including all Apache-sponsored libraries, organized by group (\texttt{logging.log4j}) and artifact (\texttt{log4j-core}). Each group-artifact pair has a series of releases, which would be downloaded and analyzed.

\vspace{.5em}
\noindent\textcircled{2} \textbf{National Vulnerability Database}\\
The National Vulnerability Database (NVD) is a centralized database maintained by the National Institute of Standards and Technology (NIST). The NVD is an internationally recognized repository of CVE and vulnerability data and utilized by numerous vulnerability researchers and widely accepted as a definite authority on recording vulnerabilities. For these reasons, we utilized the NVD to retrieve metadata about CVEs we collected.

\vspace{.5em}
\noindent\textcircled{3} \textbf{MITRE}\\
The MITRE corporation is a federally funded non-profit that addresses critical areas of cybersecurity%\cite{}
. MITRE is also the authoritative source that facilitates community development, maintains, and publishes CWEs. For this reason, MITRE was referenced to get additional data about CWEs.\\

The final source was used indirectly via the vulnerability scanner grype. The usage of grype is further elaborated in section \ref{sec:data-collection}. However, it utilizes a custom vulnerability database that aggregates vulnerability advisories from several sources, including the NVD and GitHub Security Advisories (GHSAs)\footnote{https://github.com/advisories}. The grype vulnerability database is routinely updated; therefore, to maintain consistency in our data collection, a snapshot of the grype vulnerability database was used to ensure no new CVEs were added during the extended duration of the data collection.

\subsection{Data Collection}
Data collection was done in three stages, with each stage utilizing one of the aforementioned data sources.
\label{sec:data-collection}
\begin{figure}[H]
    \centering
    \includegraphics[width=1\linewidth]{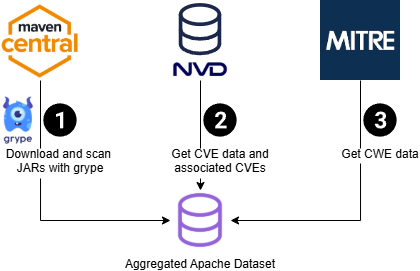}
    \caption{Data Collection Pipeline}
    \label{fig:data-collection-pipeline}
\end{figure}

\noindent\textbf{Stage 1: CVE Identification  }\\
In this stage, we developed \textbf{Graven}\cite{graven}: a Maven web crawler capable of downloading Java archives (JARs) and scanning them for CVEs using grype. Grype is a light-weight and  popular open-source vulnerability scanner\cite{grype}, capable of scanning JAR files at scale. Within a dockerized environment, Graven recursively crawled Maven Central to find URLs to valid JAR files. Releases in Maven Central often contain additional JAR files containing source Java files, documentation, or unit tests not strictly required for the library to function.
\begin{figure}[H]
    \centering
    \includegraphics[width=1\linewidth]{figures/kafka-server-excerpt.png}
    \caption{Example \texttt{kafka-server} library containing supplemental `javadoc', `sources', `test-sources', and `test' JARs}
    \label{fig:kafka-sup-jars}
\end{figure}
These JARs often end with suffixes such as `javadoc' or `sources' as seen in Figure \ref{fig:kafka-sup-jars}. In an effort to prevent duplicate CVEs for the same library and from making superfluous requests, the following JARs were omitted if they ended with any of the following suffixes [excluding the \texttt{.jar} file extension]: ``sources", ``javadoc", ``javadocs", ``tests", ``with-dependencies", ``shaded", ``minimal", ``all", ``android", ``native", ``no-deps", ``bin", ``api", ``sp", ``full". This list was obtained by analysis of repeated suffixes in the Apache ecosystem, however due to the nature of these suffixes following a convention and not a standard, additional custom suffixes may have been included as a valid JAR.

Once a valid JAR had been identified, the publish timestamp was recorded and subsequently downloaded inside the container, then scanned using grype for CVEs. The resulting grype report was processed for exclusively CVEs, additional advisories such as GHSAs or RedHat Security Advisories (RHSAs) without a CVE alias were omitted. This was done to ensure the vulnerable details could be found in the NVD in the next stage. Once scanned, the JAR was removed and the process was repeated until the entire Apache ecosystem had been processed.\\

\noindent\textbf{Stage 2: CVE Saturation \& CWE Identification}\\
% What data did we get from NVD
Following the CVE identification, we utilized the NVD API to populate the dataset with CVE CVSS scores, published timestamps, and descriptions. This stage also identified any CWEs associated with the CVE and recorded the CWE ID and related CVE to the dataset.

\noindent\textbf{Stage 3: CWE Saturation}\\
% What data did we get from MITRE
In the final stage, we used the list of CWE identifiers generated from the previous step to scrape the names of the CWEs and their descriptions from MITRE's CWE website.

\section{Results}
This section presents the findings related to the research questions. After applying our graven tool, \textbf{\nLibs\ Apache libraries} and \textbf{\nJars\ JARs} were processed. The libraries were published as early as 2005, with the most recent being from 2025. A total of \textbf{\nCVEs\ CVEs} were found, ranging in issue dates between 2007 and 2025, with \textbf{\nHits\ instances} within our dataset. The CVEs were associated with a total of \textbf{\nCWEs\ CWEs}.

% \textbf{Natural Language Processing}\\
% The National Vulnerability Database (NVD) describes CVEs. Python's NLTK library converts each description to normalized textual form by removing punctuation, making it lowercase, discarding English stop words (``to", ``the", ``of"), and applying lemmatization (reducing words to their base forms).  This ensures that trivial differences in case, punctuation, or grammatical form do not distort the analysis.

% Having processed the data, the Python script proceeds to n-gram construction and frequency distribution. Rather than focusing on single words (unigrams), it focuses on bigrams (pairs of consecutive words) and trigrams (triplets). This approach helps identify commonly used words that may reveal patterns or repeated phrases in the NVD descriptions, especially ones that indicate security themes or exploit mechanics. An additional filtering step is used to remove ``Apache" from the analysis because this study is based on the Apache ecosystem, and any reference to the term would dominate the results.

% The n-grams are grouped by year, and their frequencies are tallied. By sorting and studying the most frequent n-grams overall and by year, it is possible to highlight the most prominent word occurrences in each segment.  This method of using a macro approach for overall common sentiment descriptions and a micro approach on a year-by-year basis gives a holistic approach to recurring vulnerability patterns in the dataset.\\

\subsection{\textbf{RQ 1:} What are the persistent and repeated exploits (CWEs) that appear throughout versions?}

\begin{figure}[H]
    \centering
    \includegraphics[width=0.8\linewidth]{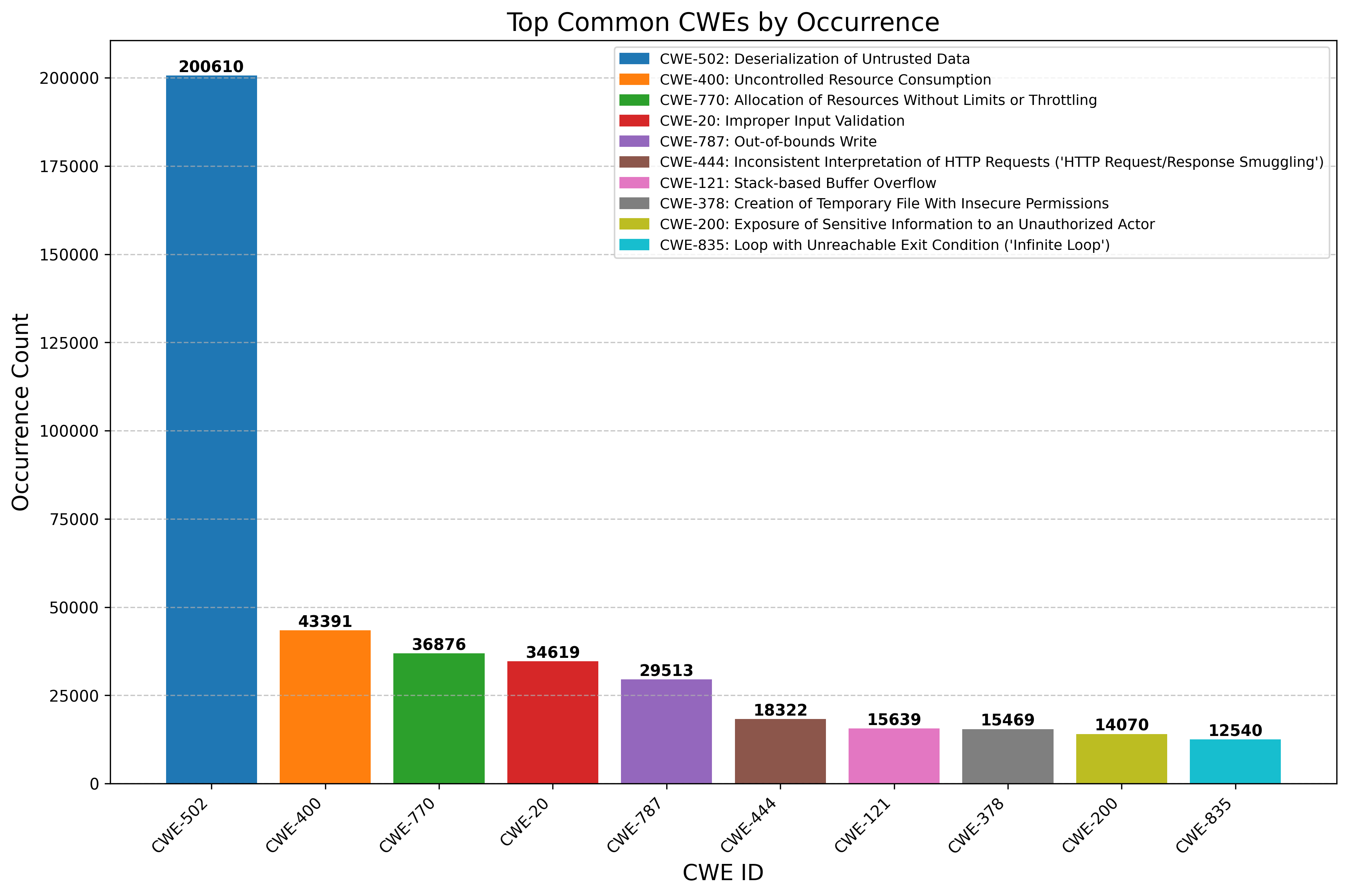}
    \caption{Most Instances of CWEs}
    \label{fig:rq1-1}
\end{figure}

To identify recurring software vulnerabilities in the Apache ecosystem, we analyzed the CVEs identified in 24,285 Apache libraries and added up their respective CWE classifications. Figure~\ref{fig:rq1-1} illustrates the top 10 most frequently occurring CWEs in our dataset.

The results shows CWE-502 (Deserialization of Untrusted Data) is responsible for over 200,000 cases—nearly five times as many as the second most common weakness. The CWE has strong risk potential as it allows for arbitrary code execution when deserializing attacker-controlled input without appropriate validation. Its commonality suggests systematic exploitation of serialization in Java-based applications, a trend of misuse in most Apache libraries. Following this, CWE-400 (Uncontrolled Resource Consumption) and CWE-770 (Allocation of Resources Without Limits or Throttling) are the next most frequent, both are related resource weaknesses that can potentially lead to denial-of-service conditions when input or system utilization is not adequately bounded.

CWE-20 (Improper Input Validation) is a common and extremely ancient family of vulnerability that appears frequently too, which implies that developers are still having difficulty in enforcing stricter input sanitation rules. Other high-frequency CWEs include CWE-787 (Out-of-bounds Write) and are typically memory corruption issues/bugs; CWE-444 (Inconsistent Interpretation of HTTP Requests), the basis of request smuggling vulnerabilities because of variations in how components handle HTTP data; and CWE-121 (Stack-based Buffer Overflow), an older but ongoing vulnerability in lower-level Apache code-bases. These CWEs are followed by CWE-378 (Creation of Temporary File with Insecure Permissions), CWE-200 (Exposure of Sensitive Information to an Unauthorized Actor), and CWE-835 (Loop with Unreachable Exit Condition), which all point towards system-level resource, and program control logic mishandling.

The frequency of these CWEs, specifically CWE 502, suggests frequent misuse of certain architectural designs and coding habits by different Apache projects. The frequency of input validation, resource management, and serialization flaws shows that there should be more stringent development rules and automated tools to detect such patterns in the initial phases of the software development cycle.

\begin{tcolorbox}[top=0.5pt,bottom=0.5pt,left=1pt,right=1pt]
\textbf{Summary for RQ 1.}
CWE-502 (Untrusted Deserialization) was by far the most common weakness, followed by CWEs related to resource exhaustion, input validation, and insecure file or memory handling. These findings indicate a set of persistent and foundational weaknesses in the Apache ecosystem, with many vulnerabilities arising from systemic misuse of serialization, inconsistent input handling, and insecure resource allocation. Targeted secure coding interventions and greater tooling support are needed to break these exploit patterns across versions.
\end{tcolorbox}

\subsection{\textbf{RQ 2:} How long do CVEs persist within libraries in the Apache ecosystem?}
The median time from the CVE disclosure to the first non-vulnerable release is 117.0 days, indicating that half of the vulnerabilities were patched within approximately four months. 
\begin{figure}[H]
    \centering
    \includegraphics[width=0.75\linewidth]{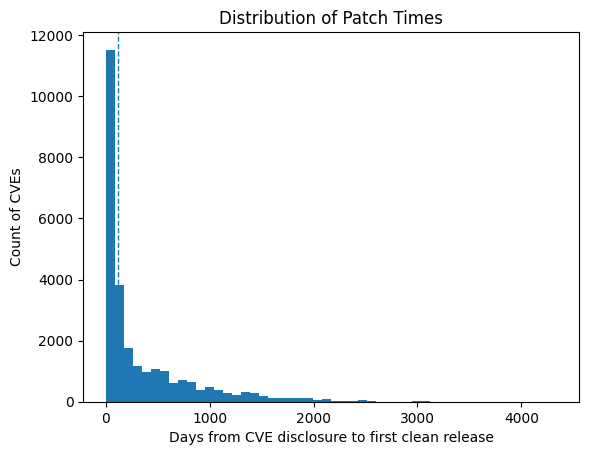}
    \caption{CVE Presence / Persistence Time Distribution}
    \label{fig:rq2-2}
\end{figure}
Figure~\ref{fig:rq2-2} presents the distribution of patch delays, showing that the vast majority of vulnerabilities are remediated in fewer than 1,000 days. Notably, extreme delays such as 4,339 days for CVE-2011-4461\cite{nistCve20114461} (library: org.apache.storm:storm-hive-examples) typically reflect anomalies in maintenance or library deprecation. The 4,339 days suggest a low prioritization or overlooked maintenance due to it being classified as a cryptography issue under CWE-310. The CWE related to CVE-2011-4461 references cryptographic issues. The shortest patch was 0 days ( CVE-2023-34055 on Library: org.apache.skywalking:apm-webapp).  This CVE with a patch of 0 days means the fix was immediately available after disclosure, which suggests rapid response and immediate vulnerability handling is effective. 

\begin{figure}[H]
    \centering
    \includegraphics[width=0.75\linewidth]{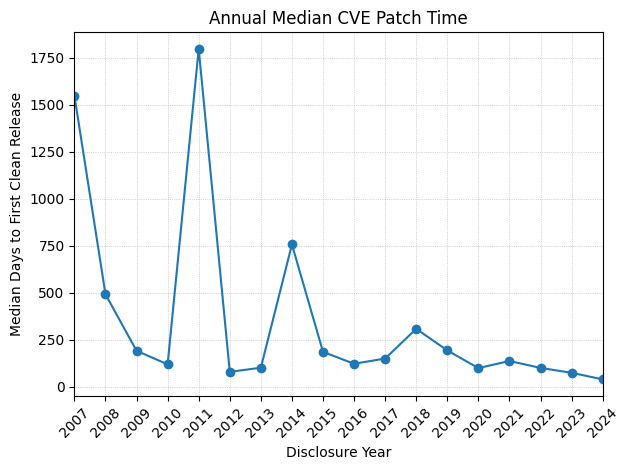}
    \caption{Annual Median CVE Patch Time}
    \label{fig:rq2-1}
\end{figure}

Figure~\ref{fig:rq2-1} shows the annual median patch time for CVEs, measured from disclosure to the first non-vulnerable release. From 2007 through 2011, medians peaked at about 1,500 days in 2007 and nearly 1,800 days in 2011. Since 2015, however, the yearly median has stabilized between roughly 100 and 300 days, suggesting that libraries today are patched more consistently and regularly. 

\begin{tcolorbox}[]
\textbf{Summary for RQ 2.}
The median time from CVE disclosure to the first non-vulnerability release is 117 days, indicating that half of the vulnerabilities are patched within four months. The overall distribution is fixed within 1,000 days, but a small trail of outliers lingers for several years. Looking at the trends over time, patch times have improved, specifically comparing 2007-2011 and 2015-now. This suggests that libraries are being patched quickly and more consistently. 
\end{tcolorbox}

\subsection{\textbf{RQ 3:} How long is the duration between the first instance of a CVE in an Apache library to its official disclosure in NVD?}
To answer this question, we compared the earliest appearance of each CVE in Apache libraries to its official publication date in the National Vulnerability Database (NVD). We found that on average, the disclosure time was 2,104 days. The longest observed delay was 6,752. Interestingly, the quickest exhibited a negative disclosure time of -3,739 days, indicating that the vulnerability was publicly disclosed long before the JAR was made available. The shortest non-negative was unsurprisingly 0 days, having the library published and CVE disclosed within the same day. 

Both the maximum and minimum values were associated with the same library version, org.apache.cocoon version 1.0.0-M1, a pre-release milestone. This suggests that certain versions, particularly early-stage releases, may exhibit greater variability in disclosure timing.
\begin{figure}
    \centering
    \includegraphics[width=0.75\linewidth]{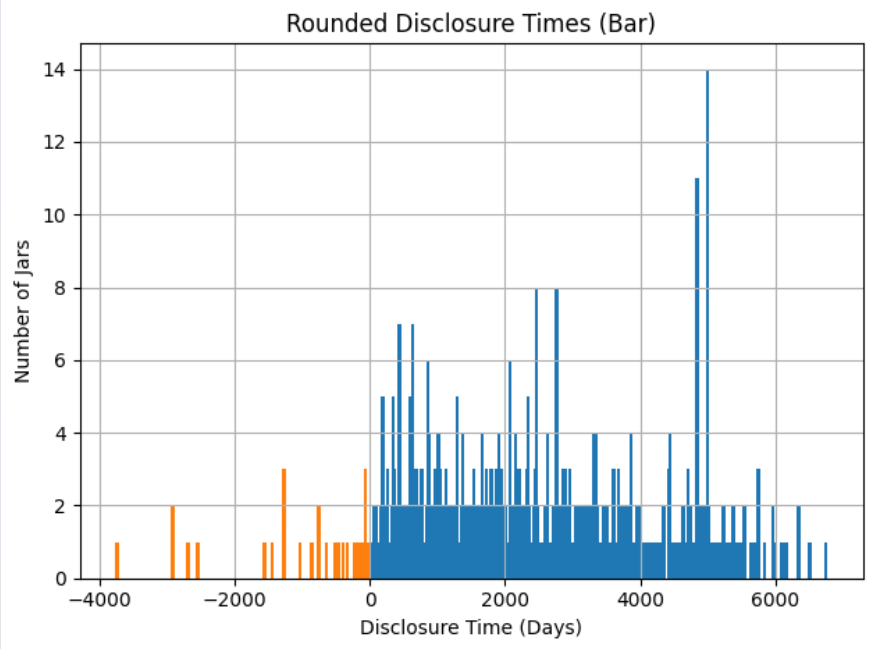}
    \caption{Average Disclosure Time}
    \label{fig:enter-label}
\end{figure}
Figure~\ref{fig:rq3-1} illustrates the distribution of disclosure times and the corresponding number of JAR files associated with each time span. As you can see, there is a noticeable amount of negative disclosure times but, the majority of the set falls between 0 and 4000 days.

\begin{tcolorbox}[top=0.5pt,bottom=0.5pt,left=1pt,right=1pt]
\textbf{Summary for RQ 3.}
Disclosure times between the earliest appearance of a CVE in an Apache library and its official publication in the NVD vary widely. While most CVEs were disclosed within 0 to 4,000 days, we observed notable outliers, including both significantly delayed and early disclosures. These results suggest variability in how and when vulnerabilities are recorded and published across different library versions.  
\end{tcolorbox}

\subsection{\textbf{RQ 4:} How long is the duration between the official disclosure of a CVE in NVD and a released version of an Apache library where the CVE has been remediated?}

% \begin{tcolorbox}[top=0.5pt,bottom=0.5pt,left=1pt,right=1pt]
% \textbf{Summary for RQ4.}
% \begin{figure}[H]
%     \centering
%     \includegraphics[width=0.8\linewidth]{figures/top_common_cwes.png}
%     \caption{Most Instances of CWEs}
%     \label{fig:rq2-2}
% \end{figure}

To address this question, we analyzed the time elapsed between a CVE's disclosure in the National Vulnerability Database (NVD) and the first version of the affected Apache library in which the vulnerability non longer appeared. This measurement, referred to as CVE Remediation Time, is a key indicator of how quickly Apache projects respond to publicly known threats.

Across 1,285 distinct CVEs tracked in our dataset, we observed a median remediation time of 99 days, suggesting that Apache library maintainers typically issue a patch within about three months following public disclosure. While this response window reflects moderate diligence in vulnerability management, it leave an exploitable window of risk for consumers, particularly when downstream projects delay updates.

\begin{figure}[H]
    \centering
    \includegraphics[width=0.75\linewidth]{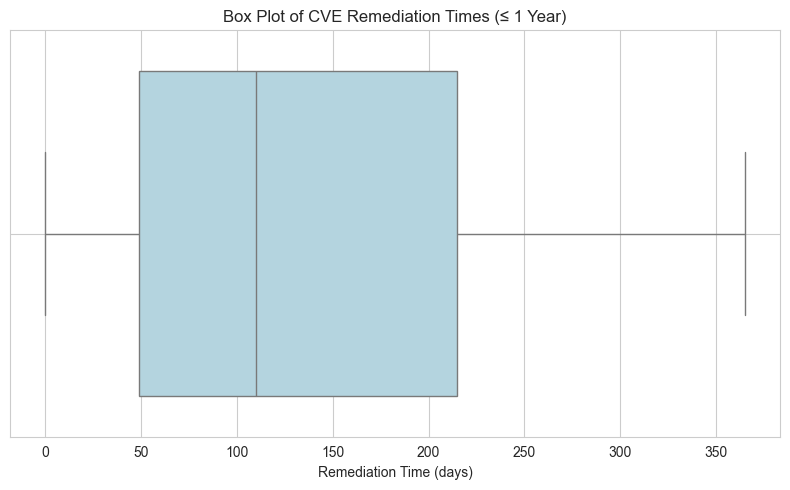}
    \caption{CVE Remediation times less than 1 Year}
    \label{fig:rq4-2}
\end{figure}

Despite this median, a significant number of CVEs take much longer to remediate. A secondary distribution analysis focusing on long-tail vulnerabilities (greater than 365 days) reveals numerous cases with remediation delays extending past 1,000 or even 3,000 days.

\begin{figure}[H]
    \centering
    \includegraphics[width=0.75\linewidth]{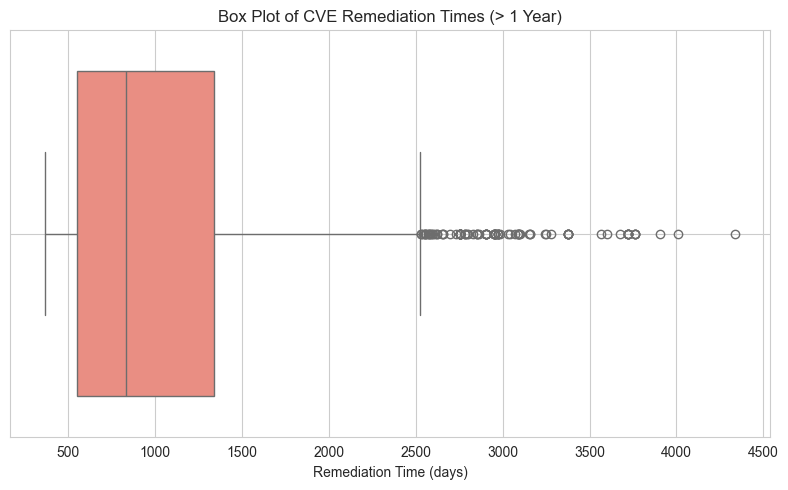}
    \caption{CVE Remediation times greater than 1 Year}
    \label{fig:rq4-1}
\end{figure}

To provide further clarity, we separated the dataset into two groups: CVEs remediated with 365 days, and those beyond. Figure 9 shows the binned CVEs with time of less than one year.

\begin{figure}[H]
    \centering
    \includegraphics[width=0.75\linewidth]{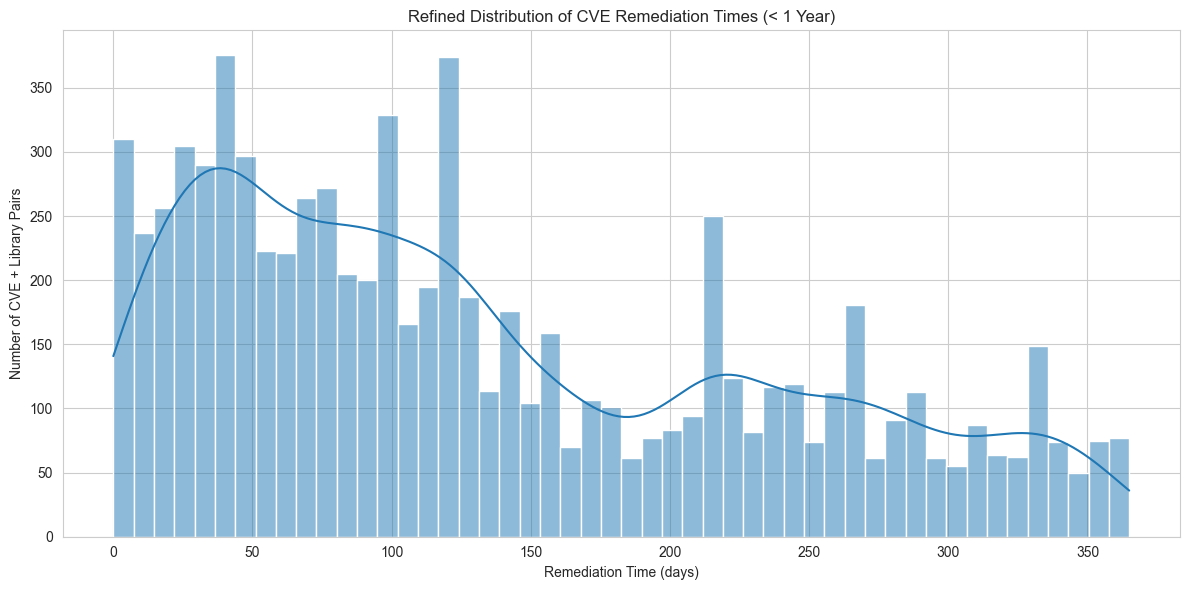}
    \caption{Distribution of CVE Remediation times less than 1 Year}
    \label{fig:rq4-4}
\end{figure}

This skewed distribution is contrasted by Figure 10, which shows the count of CVEs binned by remediation time for cases taking longer than one year.

\begin{figure}[H]
    \centering
    \includegraphics[width=0.75\linewidth]{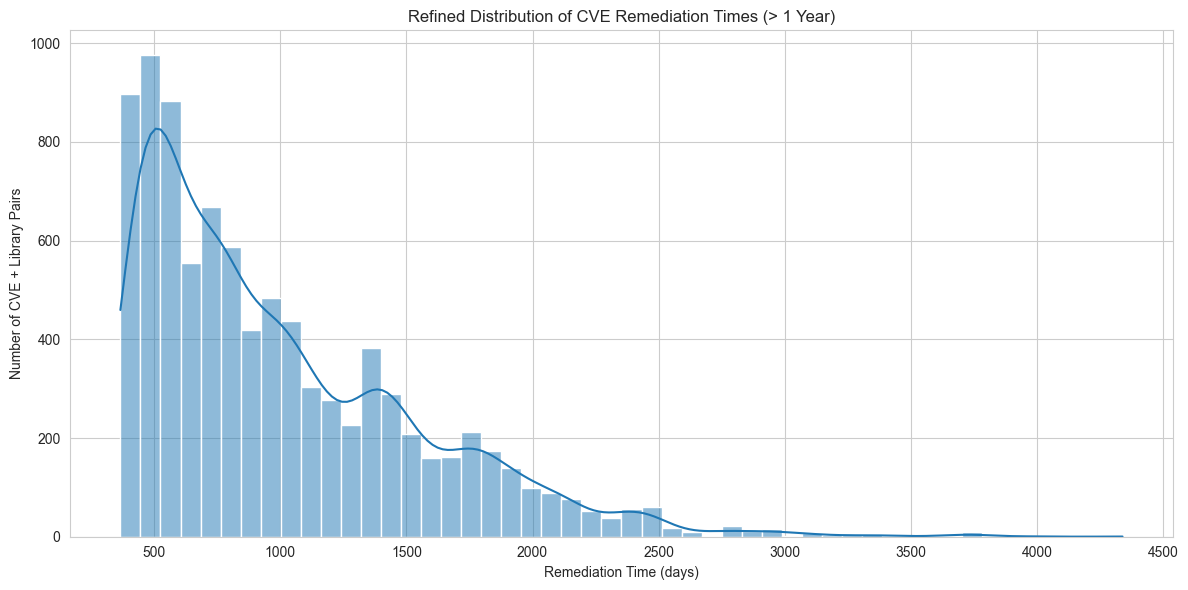}
    \caption{Distribution of CVE Remediation times greater than 1 Year}
    \label{fig:rq4-3}
\end{figure}

The causes behind these extensive delays vary. Some CVEs may involve architectural complexity, coordination across dependent components, or even disclosure timing strategies like "silent patching" or embargoed publication. In a few rare cases, early patches precede public disclosure, resulting in negative remediation time calculations. Year-over-year analysis also reveals fluctuations in average remediation time. While some years (e.g., 2011, 2015) show higher mean values exceeding 200 days, more recent years have largely stabilized between 100-150 days.

\begin{figure}[H]
    \centering
    \includegraphics[width=0.75\linewidth]{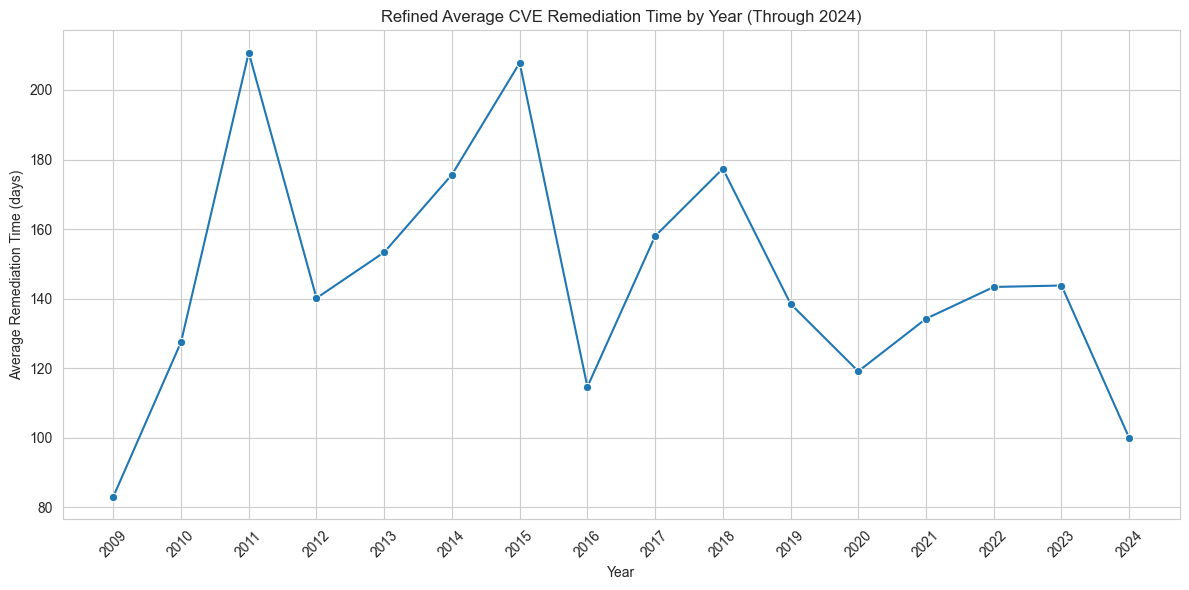}
    \caption{Average CVE Remediation time by year}
    \label{fig:rq4-5}
\end{figure}

Notably, 2008 displays a low average remediation time of just 83 days. However, this early outlier is likely not reflective of superior patching practices but rather an artifact of limited CVE detection an documentation capabilities during that period. Earlier in the dataset, many vulnerabilities simply went unreported or undisclosed due to less mature tooling and weaker security transparency norms. As the CVE ecosystem improved with better tooling like Grype and broader community participation, reporting improved and has led to a more representative distribution of remediation delays.

The combination of statistical median, distribution plots, and yearly averages offers a nuanced view: although many CVEs are addressed in a timely manner, a persistent subset suffers from remediation delays spanning several years. These delays stem from transitive inclusion of affected packages (e.g., Jackson Databind, Guava), which are used across thousands of downstream libraries. Such findings align with broader software supply chain security concerns and reinforce calls for better automation in vulnerability tracking.

\begin{tcolorbox}[top=0.5pt,bottom=0.5pt,left=1pt,right=1pt]
\textbf{Summary for RQ 4.}
CVE remediation times across the Apache ecosystem show moderate industry alignment with patch time expectations (e.g., ~90 days). However, the presence of long-tail delays highlights the continued importance of proactive dependency monitoring, timely updates, and enhanced coordination across project maintainers.
\end{tcolorbox}

% \end{tcolorbox}

\section{Discussion}

The ten most recurring CVEs included CVE-2020-8908, CVE-2023-2976, CVE-2022-42004, CVE-2022-42003, and CVE-2018-10237 along with other vulnerabilities from 2020 and 2024. Each of these vulnerabilities were associated with more than 500 distinct artifact identifiers, indicating substantial reach across dependency chains. While CVE-2021-44228 (Log4Shell) is not among the most frequent in this specific dataset, the top entries appear to reflect recent vulnerabilities that are either embedded across many packages or represent reused patterns in code across multiple components. This highlights that high severity does not necessarily correlate with ubiquity.

The nature of these CVEs, while diverse in their exact specifications, commonly suggests the presence of generalizable exploit conditions or widely reused vulnerable libraries. For instance, CVE-2020-8908 pertains to a temporary directory creation vulnerability in all versions of Google Guava, allowing potential unauthorized access to data in temporary directories\cite{nistCve20208908}. Similarly, CVE-2023-2976 involves the use of Java's default temporary directory for file creation in Guava versions 1.0 to 31.1, which could allow other users and applications to access those files\cite{nistCve20232976}. CVE-2022-42004 and CVE-2022-42003 are vulnerabilities in the FasterXML Jackson Databind library that appeared prominently in the CVE description analysis, both involving improper deserialization handling that can lead to resource exhaustion through deeply nested input structures\cite{nistCve202242004}\cite{nistCve202242003}.CVE-2018-10237, meanwhile, allows remote attackers to conduct denial of service attacks through Google Guava\cite{nistCve201810237}. These examples underscore how vulnerabilities in foundational packages can silently affect downstream systems through transitive inclusion.

In several cases, CVEs exhibited a substantial delay between initial discovery and official publication in the NVD. These delays may arise from coordinated disclosure practices, where vendors are given time to patch and communicate before public release. Additionally, complex vulnerabilities that require extensive internal validation or involve multiple parties could take longer to triage and document formally. In certain situations, organizational policy or national-level database approval procedures may introduce bureaucratic delays. These factors contribute to inconsistent timelines, even for vulnerabilities with clear and high-impact implications.

Trend analysis from 2007 through 2024 reveals a pronounced acceleration in the rate of CVE publication beginning in 2016, as confirmed by year-by-year frequency data. The most significant surges occurring in the aftermath of high-profile cybersecurity incidents. In particular, 2017 saw a notable increase in disclosures following the exploitation of CVE-2017-5638, which contributed to the Equifax data breach\cite{equifaxEquifaxReleases}. Similarly, 2021 witnessed a second spike driven by the global response to Log4Shell (CVE-2021-44228), a critical remote code execution vulnerability in Apache Log4j\cite{cybersecuritydiveYearsLog4j}. These patterns underscore the increasing importance of transparent vulnerability reporting and the expanding focus on supply chain security throughout the software industry.
\section{Threats to Validity}

This section reviews potential challenges to our findings.

\subsection{Tracking Sequential Versioning}
Consistent versioning is a challenge to all software, with developers often having their own preferred system and caveats, such as revisions or alpha and beta releases. While sematic versioning is the most widely adopted version, this too often sees slight variations from the official standard. These deviations make it increasingly challenging to determine the order in which software developed that is consistent across all projects. We determined the most reliable way to determine a sequential package is to find the next most immediate timestamp as logically the next release follows the current. However, this does not account for parallel or non-orthodox release times, resulting in interwoven timelines that are in relativity isolated from each other. This could create false positive cases where timelines are not completely representative of the actual library development time.

\subsection{Jar Exclusion}
Library releases often contain jars containing source code, documentation, tests, and other related content. Due to time and size constraints, we omitted several jars as they were not utilized by build systems like Maven and Gradle when compiling code, however there may be additional CVEs and CWEs that would go unnoticed due to their omission. Additionally, these jars may also serve a functional purpose, such as architecture specific distributions, that would be more likely to be used downstream and contain a previously undisclosed CVEs.

% \section{Future Work}

\section{Conclusion}
In this paper we examined the history of CVE presence, disclosure, and patching rates within the Apache ecosystem. We analyzed \nJars\ jars spanning across \nLibs\ Apache Libraries and provided empirical evidence of the state of vulnerability management. In total, we found \nCVEs\ CVEs and \nCWEs\ associated CWEs with the Apache Libraries, with an median disclosure time of 1894 days, remediation time of 99 days, and overall presence of 536 days.

Utilizing our graven tool, future work can expand to other ecosystems or Maven Central in its entirety. A larger scope would provide a more broader insight into CVE rates, as well as highlight any particularly frequent CWEs.

\bibliographystyle{plain}
\bibliography{references, ref}

\end{document}